\documentclass[fleqn,twoside]{article}
\usepackage[centertags]{amsmath}
\usepackage{espcrc2}
\usepackage{graphicx}
\usepackage[figuresright]{rotating}
\usepackage{tabularx}

\newcommand{\be}{\begin{equation}}
\newcommand{\ee}{\end{equation}}

\title{Investigations on the  deconfining phase transition in QCD}

\author{P. Cea\address{Dept. of Physics  and INFN - Bari - Italy},
        L. Cosmai\address{INFN - Bari - Italy}, and
        M. D'Elia\address{Dept. of Physics and INFN - Genova - Italy}}

\begin{document}

\begin{abstract}
We investigate
the  deconfining phase transition in SU(3) pure gauge theory and in full QCD
with two flavors of staggered fermions  by means of a gauge invariant 
thermal partition functional.
In the pure gauge case our finite size scaling analysis 
is in agreement with the well known 
weak first order phase transition.
In the case of 2 flavors full QCD we find that the phase
transition is consistent  with
weak first order, contrary to the expectation of a 
crossover for not too large quark masses.
\vspace{1pc}
\end{abstract}

\maketitle

\section{INTRODUCTION}

To detect the deconfinement phase transition in pure gauge theories 
the expectation value
of the trace of the Polyakov loop is used as an order parameter. 
In presence of dynamical fermions the  Polyakov loop ceases 
to be an order parameter since Z(N)
symmetry is no longer a symmetry of the action.
Alternatively deconfinement could be detected by looking at
Abelian monopole condensation which 
can be detected~\cite{DiGiacomo:1999fa}
by means of the vacuum expectation value of a magnetically charged operator
 whose expectation value is different from zero in the confined phase and goes 
to zero at the deconfining phase transition.

In a different way, abelian monopole condensation 
can be signaled by the free energy needed to create an abelian monopole
in the vacuum. This free energy can be computed~\cite{Cea:2000zr,Cea:2001an}
through a  gauge invariant lattice effective action which in turn 
is defined at zero temperature by means of the lattice Schr\"odinger functional 
and at finite temperature by means of a thermal partition functional.
Since abelian monopole condensation 
is not related,
as the Polyakov loop, to a symmetry of the lattice action, 
it could still be useful in order to detect a phase transition 
even if dynamical fermions are included into the action.
The lattice thermal partition
functional in presence of a static background field
is defined~\cite{Cea:2001an,Cea:2002wx} as
\be
\label{ZetaTnew} \mathcal{Z}_T \left[ \vec{A}^{\text{ext}} \right]
= \int_{U_k(L_t,\vec{x})=U_k(0,\vec{x})=U^{\text{ext}}_k(\vec{x})}
\mathcal{D}U \, e^{-S_W}   \,,
\end{equation}
$T=1/a L_t$ is the physical temperature. T
he spatial links belonging
to the time slice $x_t=0$ are constrained to the value of the external background field,
\be
\label{coldwall}
U_k(x)|_{x_t=0} = U^{\text{ext}}_k(\vec{x})
\,,\,\,\,\,\, (k=1,2,3) \,\,,
\end{equation}
$U^{\text{ext}}_k(x)$ being the lattice version of
the external continuum gauge field,
the temporal links are not constrained.
At finite temperature the relevant quantity is the
free energy functional:
\be
\label{freeenergy} 
\frac{F[\vec{A}^{\text{ext}}]}{T_{\text{phys}}}= 
-  \ln \frac{\mathcal{Z}_T[\vec{A}^{\text{ext}}]} {{\mathcal{Z}}_T[0]}
\,.
\end{equation}
When including dynamical fermions, the thermal partition functional
in presence of a static external background gauge field, Eq.~(\ref{ZetaTnew}),
becomes:
\be
\label{ZetaTfermions}
\begin{split}
& \mathcal{Z}_T \left[ \vec{A}^{\text{ext}} \right]  = 
\nonumber \\ 
& \quad =  \int_{U_k(L_t,\vec{x})=U_k(0,\vec{x})=U^{\text{ext}}_k(\vec{x})}
\mathcal{D}U e^{-S_W} \, \det M \,,
\end{split}
\end{equation}
where $S_W$ is the Wilson action
and $M$ is the fermionic matrix.
Notice that the fermionic fields are not constrained and
the integration constraint is only relative to the gauge fields: 
this leads, as in the usual QCD partition function, to the appearance of 
the gauge invariant fermionic determinant after integration on the 
fermionic fields. 

To detect  monopole  condensation we consider~\cite{Cea:2000zr,Cea:2000rj} the 
following quantity defined in terms of  free energy $F$ needed to create a monopole
in the vacuum:
\be
\label{disorderT}
e^{-F/T_{\text{phys}}} = \frac{\mathcal{Z}_T
\left[ \vec{A}^{\text{ext}}  \right]} {\mathcal{Z}_T[0]} \,.
\end{equation}
In presence of monopole  condensation $F$ is finite
 and $e^{-F/T_{\text{phys}}} \ne 0$.
In practice it is easier to compute $F^\prime(\beta)$ the $\beta$-derivative   of the free energy
(eventually, since $F=0$ at $\beta=0$, $F$  can be obtained by numerical integration of
$F^\prime(\beta)$).

\section{SU(3) PURE GAUGE}

We  consider SU(3) pure gauge in an abelian monopole background 
field. The gauge potential in the continuum is
\be
\label{monop3su2}
g \vec{b}^a({\vec{x}}) = \delta^{a,3} \frac{n_{\mathrm{mon}}}{2}
\frac{ \vec{x} \times \vec{n}}{|\vec{x}|(|\vec{x}| -
\vec{x}\cdot\vec{n})} \,,
\end{equation}
where $\vec{n}$ is the direction of the Dirac string and $n_{\text{mon}}$ (integer)
is the number of monopoles.
On the lattice the spatial links exiting from the sites at the boundary 
of the time slice $x_t=0$ are constrained to the (lattice version of) the
gauge potential Eq.~(\ref{monop3su2}).
Links at spatial boundaries ($x_t \ne 0$) can be constrained to the
monopole background field (``spatial fixed boundary conditions'') or can
be periodic (``spatial periodic boundary conditions'').
These two different choices (for the monopole field
which vanishes at infinity) are equivalent in the thermodynamical limit.
The simulations were performed on lattices of different spatial sizes 
($16^3$, $24^3$, and  $32^3$) and 
fixed the temporal extent ($L_t=4$), using APEmille/crate in Bari. 
We find that our data for $F^{\prime}(\beta,L_s^{\text{eff}})$ can be fitted according to 
the following scaling law
\be
\label{scalinglaw}
F^{\prime}(\beta,L_s^{\text{eff}}) = 
\frac{a_1 (L_s^{\text{eff}})^{\gamma}}{\left|  (L_s^{\text{eff}})^{1/\nu} (\beta - \beta_c) - d_1 \right|^\alpha} \,,
\end{equation}
where the above scaling relation holds quite well 
for a large range of the scaling variable $x =  (L_s^{\text{eff}})^{1/\nu} (\beta - \beta_c)$, 
with  $\nu = 0.334 \pm 0.021$  consistent with a first order phase (see Fig.~1).
%
%
%
\begin{figure}[!ht]
\label{cazzo}
\begin{center}
\includegraphics[width=0.5\textwidth,clip]{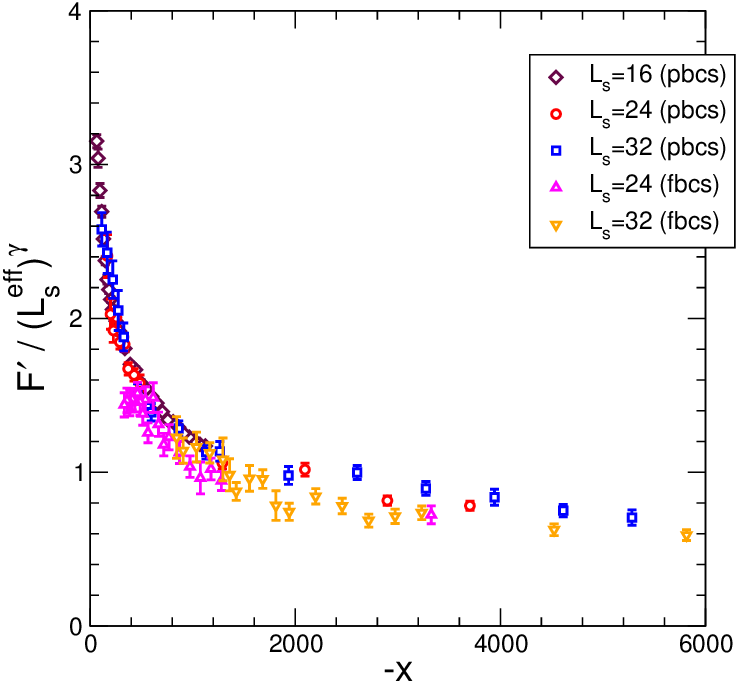}
\vspace{-1.1cm}
\caption{We plot together data  for $F^{\prime}$ 
(fbcs: ``spatial fixed boundary conditions'', $L_s^{\text{eff}}=L_s-2$)  
and (pbcs: ``spatial periodic boundary conditions'', $L_s^{\text{eff}}=L_s$).
}
\end{center}
\end{figure}
%
%
%
%
%
%
%


\vspace{-1.2cm}
\section{QCD ($N_f=2$)}

We study  QCD with two dynamical staggered fermions.
We simulate the theory with  a ``cold'' time-slice  where (as in the case
of pure gauge) the spatial links are constrained to the abelian monopole background 
field. 
%
%
%
\begin{figure}[!ht]
\label{Fig2}
\includegraphics[width=0.5\textwidth,clip]{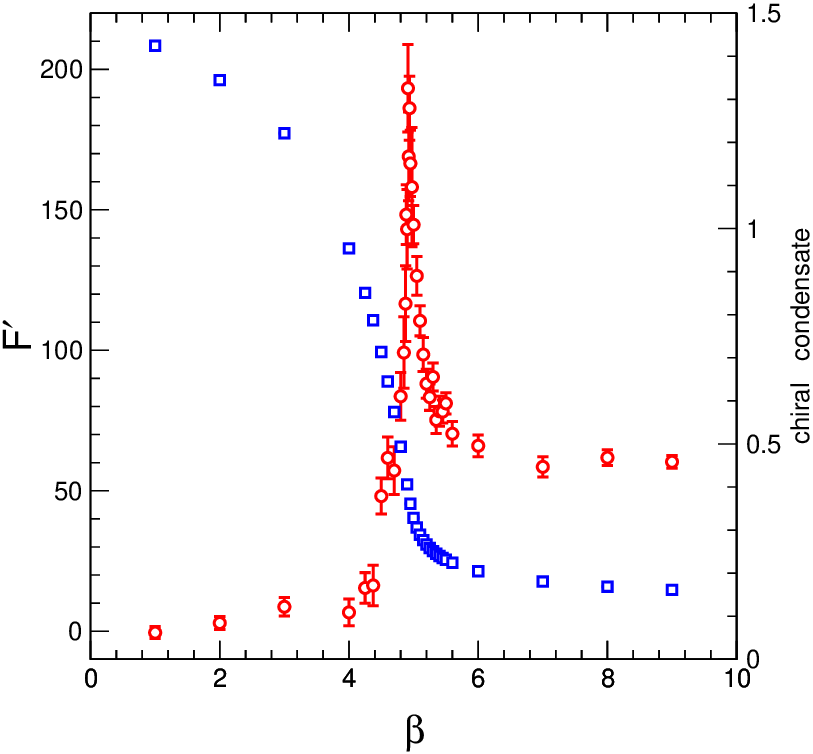}
\vspace{-1.1cm}
\caption{
$F^{\prime}$  on a $16^3\times4$ lattice in the case of full QCD with 2 dynamical flavors, is displayed
together with chiral condensate.}
\end{figure}
In Fig.~(\ref{Fig2}) the derivative of the free energy with respect to
the gauge coupling $F^{\prime}(\beta)$  is displayed
together with chiral condensate data  suggest that the peak in $F^{\prime}(\beta)$
corresponds to the drop of the chiral condensate.
Using our data for $F^{\prime}(\beta)$ on different spatial volumes
and different bare quark masses we try to infer the critical behavior of two flavors 
full QCD  near the phase transition.
We varied the lattice size $L_s$ ($L_s=16,20,32$)  
and the staggered quark mass $m_q$ ($m_q=0.075,0.2676,0.5003$). 
At fixed $m_q$, as in the quenched
case, the peak increases as the lattice volume is increased.
At fixed spatial volume the critical coupling depends on $m_q$.
All our lattice data can be described 
by a finite size scaling function where the gauge critical coupling now does depend on 
the quark mass and is determined by the chiral critical 
point\cite{Karsch:1994hm,Engels:2001bq,Karsch:2000kv}
\be
\label{scalingfermions}
\begin{split}
& F^{\prime}(\beta,L_s^{\text{eff}},m_q) = \\
& \quad
\frac{a_1 (L_s^{\text{eff}})^{\gamma}}{\left|  (L_s^{\text{eff}})^{1/\nu} (\beta - \beta_c(0) -c m_q^{\eta}) - d_1 \right|^\alpha} \,,
\end{split}
\end{equation}
the relevant chiral critical exponents being compatible with 
those of the three-dimensional O(4) symmetric
spin models~\cite{Engels:2001bq}.
We find that the critical exponent $\nu=0.31\pm0.03$ is consistent 
with a first order phase transition,  even though 
weaker than in SU(3) pure gauge (analogous indications 
have been obtained by the Pisa group~\cite{Carmona:2002ty}).
%
%
%
\begin{figure}[!ht]
\label{Fig3}
\includegraphics[width=0.5\textwidth,clip]{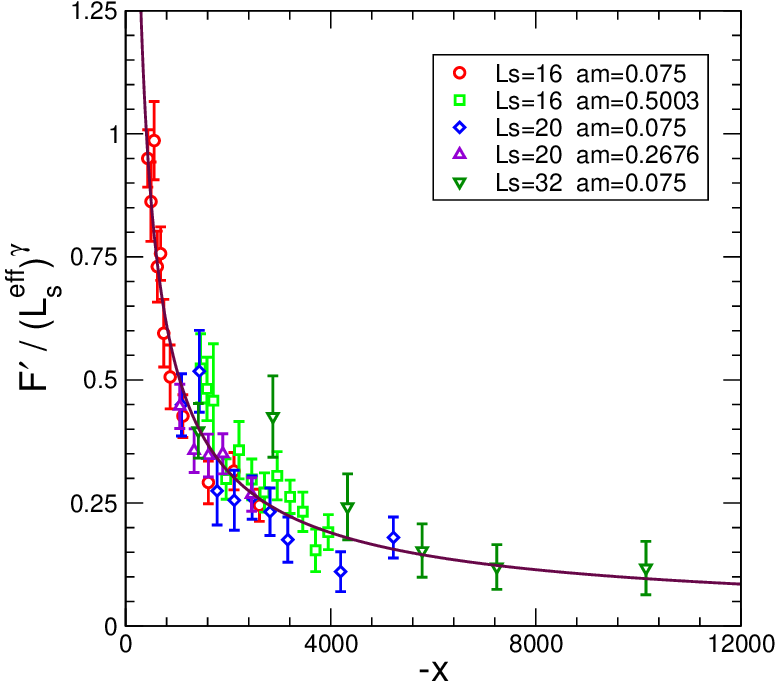}
\vspace{-1.1cm}
\caption{$F^{\prime}(\beta,L_s^{\text{eff}},m_q)$ rescaled by the factor $(L_s^{\text{eff}})^\gamma$.
The values of $L_s^{\text{eff}}$ and $m_q=am$ are displayed in the legend.}
\end{figure}


\section{CONCLUSIONS}

We investigated  the  phase transition in SU(3) pure gauge theory and in full QCD
with two flavors of degenerate staggered fermions.
To locate the phase transition
we looked at the free energy to create a monopole in the vacuum.
In the pure gauge case our finite size scaling analysis is consistent with the well known
weak first order phase transition. 
In the case of QCD  we find that the phase 
transition is compatible  with
a weak first order phase transition (contrary to the expectation of a crossover 
for not too large quark masses).

\end{document}